\documentclass[12pt,fleqn]{article}

\usepackage{amsmath,amsthm}
\usepackage[T1]{fontenc}
\usepackage[utf8]{inputenc}
\usepackage{authblk}
\usepackage{natbib}
\usepackage{url}
\usepackage[pdftex]{graphicx}
\usepackage{times}
\usepackage{float}
\usepackage[title]{appendix}
\usepackage{epstopdf}
\usepackage{soul, color}

\newcounter{lastnote}

\newtheorem{prop}{Proposition}

\title{Valuation of a Bermudan DB Underpin hybrid pension benefit}
\author[1]{Xiaobai, Zhu\thanks{Department of Statistics and Actuarial Science (SAS), University of Waterloo, 200 University Ave. West, Waterloo, ON, Canada, N2L 3G1. Email: x32zhu@uwaterloo.ca}}

\affil[1]{Department of Statistics and Actuarial Science, University of Waterloo}
\author[1]{Mary, Hardy}
\author[1]{David, Saunders}

\begin{document}
	
	
	\baselineskip16pt
	
	
	\maketitle
	\begin{abstract}
In this paper we consider three types of embedded options in pension benefit design.\\

The first is the Florida second election (FSE) option, offered to public employees in the state of Florida in 2002. Employees were given the option to convert from a defined contribution (DC) plan to a  defined benefit (DB) plan at a time of their choosing.  The cost of the switch was assessed in terms of the  ABO (Accrued Benefit Obligation), which is the expected present value of the accrued DB pension at the time of the switch. If the ABO was greater than the DC account, the employee was required to fund the difference.\\

The second is the DB Underpin option, also known as a floor offset, under which the employee participates in a DC plan, but with a guaranteed minimum benefit based on a traditional DB formula.\\

The third option can be considered a variation on each of the first two. We remove the requirement from the FSE option for employees to fund any shortfall at the switching date. The resulting option is very similar to the DB underpin, but with the possibility of early exercise. Since we assume that exercise is only permitted at discrete, annual intervals, this option is a Bermudan variation on the DB Underpin.  \\
	
	 We adopt an arbitrage-free pricing methodology to value the option. We analyse and value the optimal switching strategy for the employee by constructing an exercise frontier, and illustrate numerically the difference between the FSE, DB Underpin and Bermudan DB Underpin options.\\
		
	\end{abstract}

	\section{Introduction}
 Over the past two decades, there has been a significant global shift in employer sponsored pension plans from Defined Benefit (DB) to Defined Contribution (DC). According to \citet{pensionasset2017}, DC pension assets in 22 major pension markets have grown from 41.1\% in 2006 to 48.4\% in 2016. However, there is also some growing recognition that DC plans may not always be the best option. In 2013 the OECD\footnote{\url{http://www.oecd.org/finance/private-pensions/designingfundedpensionplans.htm}} identified six common flaws in DC systems, including excessive volatility of funds in the pre-retirement phase, and decumulation options which are ``\emph{not fit for purpose}''.
	
	Hybrid pension plans, which combine features of DC and the DB plans, may be able to meet employer and employee needs better than DC or DB individually. Numerous  hybrid pension structures have been proposed, with different objectives and different balance of employee and employer risk allocation. \citet{survey} and \citet{surveyHybrid} present comprehensive surveys on hybrid plans. Some popular examples include Cash Balance pension plans, Target Benefit pension plans and DB Underpin plans. Studies have  demonstrated that the hybrid pension market is expanding in some areas. For example, \citet{Kravitz} shows that the number of Cash Balance Plans in the U.S.  increased from 1337 in 2001 to 15178 in 2014).
	
	The rising interest in hybrid pension plans suggests the potential for designing new risk-sharing schemes. In this paper, a new hybrid pension design is introduced. The intuition behind the new plan is based on two existing embedded options: the second-election option in the Florida retirement system, and the DB underpin plan. This paper evaluates the new option using the market consistent valuation method (see \citet{Boyle03guaranteedannuity}, \citet{Marshall} and \citet{chen2009} for the valuation of various insurance and pension products) and we compare it with the Florida second election  and the DB underpin options. This paper illustrates how the flexible nature of hybrid pension plans allows sponsors to design new risk-sharing schemes based on their risk appetite and objectives.

	The remainder of the paper is structured as follows. Section 2 introduces the notation and assumptions, and provides detailed background information for the second-election option and the DB underpin option. Section 3 presents the problem formulation for the new pension plan, as well as some theoretical results in discrete time. Section 4 extends the work into the continuous case and incorporates stochastic salaries. Section 5 displays the numerical results. Section 6 concludes.

	\section{Notation and Assumptions}

	\begin{itemize}
		\item[$c$] denotes the annual contribution rate (as a proportion of salary) into the DC plan. We assume that contributions are paid annually. We also assume that all contributions are paid by the plan sponsor/employer, although this is easily relaxed to allow for employee contributions.
\item[$T$] denotes the time of retirement of the employee.
		\item[$r$] denotes the risk free rate of interest, compound continuously.
		\item[$b$] denotes the accrual rate in the DB plan.
		\item[$\ddot{a}(T)$] denotes the actuarial value at retirement of a pension of 1 per year.
		\item[$S_t$] denotes the stochastic price index process of the funds in the DC account. We assume $S_t$ follows a Geometric Brownian Motion, so that
		\begin{align*}
			\frac{dS_t}{S_t} = rdt + \sigma_SdZ_S^Q(t)
		\end{align*}
		where $Z_S^Q(t)$ is a standard Brownian Motion under the risk neutral measure $Q$.
		\item[ $L_t$ ]denotes the salary from $t$ to $t+1$ for $t=0,1,...,T-1$, where $t$ denotes years of service. We assume that salaries increase deterministically at a rate $\mu_L$ per year, continuously compounded, so that
		\[L_t=L_0e^{t\mu_L}\]
		We will also consider a stochastic salary process, in the continuous setting, assuming it is hedgeable through the financial market:
		\begin{align*}
		  &  \frac{dL_t}{L_t}= rdt + \sigma_LdZ_L^Q(t)\\
		  &  dZ_L^Q(t)dZ_S^Q(t)= \rho dt
		\end{align*}
		where $Z_L^Q(t)$ is a standard Brownian Motion under the risk neutral measure $Q$ and $\rho$ is the correlation coefficient between $Z_L^Q$ and $Z_S^Q$ (correlation between the stock market and the salary increase).
		\item[] We  ignore mortality and other demographic considerations.
	\end{itemize}
	
	\section{DB-Underpin Plan}
	The DB Underpin pension plan, also known as the floor-offset plan in the USA, provides a guaranteed defined benefit minimum which underpins a  DC  plan. Plan sponsors make periodic contributions into the member's DC account, and separately contribute to the fund which covers the guarantee. Employees usually have limited investment options to make the guarantee value more predictable. At the retirement date, after  $T$ years of service, if the member's DC balance is higher than the value of the  guaranteed minimum defined benefit pension, the plan sponsor will not incur any extra cost. However, if the DC benefit is smaller, the plan sponsor will cover  the difference.
	
	Using  arbitrage-free pricing, we can calculate the present value of the cumulative cost of the DB underpin plan at time $t=0$ as follows. We assume (more for clarity than necessity) that DB benefits are based on the final 1-year's salary, and we ignore all exits before retirement.

	\begin{align*}
	C^U = \underbrace{E^Q\left[\sum_{t=0}^{T-1}e^{-rt}cL_t\right]}_{\text{\small PV of DC Contributions}} + \qquad \underbrace{E^Q\left[e^{-rT}\left(bT\ddot{a}(T)L_{T-1} - \sum_{t=0}^{T-1}\frac{S_T}{S_t}cL_t\right)_{{\large +}} \right]}_{\mbox{\small Value of DB underpin option}}
	\end{align*}
	
	Notice that the option value does not have an explicit solution, but can be determined using Monte Carlo simulation. See \cite{chen2009}  for details on the valuation and funding of the DB underpin option.
	
	\section{Florida Second Election (FSE) Option}
	In 2002, public employees of the State of Florida were given an option to switch from their DC plan to a DB plan anytime before their retirement date. The cost of participating in the DB plan is calculated by the accumulated benefit obligation (ABO), which is the present value of the accrued benefit, based on current service and current pensionable salary.
	
	We denote the ABO of the DB benefit for an employee with $t$ years of service as $K_t$, so that
	\begin{align*}
	&
	K_t=b \, L_{t-1} \, t\, \ddot{a}(T) \, e^{-r(T-t)}
	\end{align*}
	
	Under the FSE hybrid plan, let $\tau$ denote the time that the employee shooses to switch from the DC plan to the DB plan.  If the ABO at $\tau$  is more than the DC account balance, the employee needs to fund  the difference from her own resources. If the DC account is more than the ABO at transition, then the employee retains the difference in a separate DC top-up account which can be withdrawn at retirement.
	
	Mathematically, the present value of the total DC and DB benefit cost at inception is
	\begin{align*}
	C^{se} = &\sup_{0\leq \tau \leq T} E^Q\left[\sum_{t=0}^{\tau-1}e^{-rt}cL_t
	+
	e^{-r\tau}\left(K_Te^{-r(T-\tau)}-K_\tau \right)\right]\\
	\end{align*}
	where the first term, as in the previous section, is the present value of the DC contributions, and the second term is the additional funding required for the DB benefit, offset by the ABO at transition, which is funded from the DC contributions. The `$\sup$' indicates that the valuation assumes the switch from DC to DB is made at the time to maximize the cost to the employer. \cite{milevsky}  studied the price and optimal switching time of the Florida option with deterministic assumptions.

	\section{Bermudan DB underpin plan - Discrete Case}
	The FSE design has the advantage that it provides  employees with the flexibility to choose their plan type based on their changing risk appetite. However, employees retain the investment risk through the DC period of membership, and also have the additional risk of a  suboptimal choice of  switching time. Moreover, when the DC investment falls  below the ABO of the DB plan, the employee may be  unable to switch, if they do not have sufficient assets to make up the difference.
	
	Inspired by the idea of combining the DB underpin plan with the Florida option, we investigate a new hybrid design, which we call the Bermudan DB underpin, which adds a guarantee at the time of the switch from DC to DB. If the employee's  DC account is below the ABO when she elects to switch, then the plan sponsor will cover the difference.
	
	\subsection{Problem Formulation }
	We assume that contributions are made annually into the DC account until the employee switches to DB, and that we are valuing the benefits at $t$.
	We assume also that the employee has not switched from DC to DB before time $t$,  that the DC account is $w_t$ at $t$, and that future account values (up to the switching time) follow the process
\begin{align*}
	W_{t+\tau} = w_t\frac{S_{t+\tau}}{S_t} + \sum_{u=t}^{\tau-1}\frac{S_{t+\tau}}{S_u} cL_u \qquad \tau=1,2,...T-t.
	\end{align*}

 Then the present value at time $t$ of the cost of future benefits (past and future service, DC and DB), which is denoted by $C_1(t,w_t)$, can be expressed as the sum of three terms:
	
	\begin{enumerate}
		\item The present value of the future contributions into the DC account before the member switches to the DB plan.
		\item The present value of the cost of the DB benefit, offset by the ABO at the time of the switch.
		\item The difference between the ABO and the DC balance at the switching time, if positive.
	\end{enumerate}
	
			We  assume that the exercise dates are at the beginning of each year, before the contribution is made into the DC account, so the admissible exercise dates are $\tau=0,1,\cdots T-t$. We take the maximum value of the sum of these three parts, maximizing over all the possible switching dates, as follows.
	\begin{align*}
	C_1(t,w) =\sup_{\tau \in [0, 1,\cdots,T-t]} E^Q_t\bigg[\sum_{u=0}^{\tau-1}e^{-ru}cL_{u}&+
	e^{-r \tau}   \left(K_Te^{-r(T-t-\tau)}-K_{t+\tau}\right) \\
	& \quad
	+ e^{- r\tau}\left(K_{t+\tau} - W_{t+\tau}\right)^{+} \bigg|W_t=w \bigg]
	\end{align*}
	Notice that the switching time $\tau$ is involved in all three parts, which makes the analysis more complex. However, using put-call parity, the problem simplifies to:
	\begin{align}\label{cost_function}
C_1(t,w) = &E^Q_t\left[K_Te^{-r(T-t)}\right] \nonumber \\
	&\quad + \sup_{\tau \in [0, 1,\cdots,T-t]}\, \!E^Q\!\left[ e^{-r\tau }\left(W_{\tau+t}-K_{t+\tau}e^{-r(T-t-\tau)}\right)^+\bigg|W_t=w \right] \nonumber\\
& \qquad \qquad   -w
	\end{align}
	Details are given in  Appendix A. The new formulation  also consists of three terms:
	\begin{itemize}
		\item The present value of the DB plan benefits at time $t$.
		\item A Bermudan type call option, with underlying process $W_t$ and strike value $K_t$.
		\item Offset by the existing DC balance at $t$.
	\end{itemize}
	The first and third terms do not depending on the switching time, and the first part does not even depend on the DC balance. To study the optimal exercising strategy, we omit the first and third terms and define our value function as
	\begin{align}
&	v_1(t,w)= \sup_{\tau \in [0, 1,\cdots,T-t]} E^Q_t\left[e^{-r\tau}\left(W_{t+\tau}-K_{t+\tau}e^{-r(T-t-\tau)}\right)^+ \bigg| W_t=w\,\right] \label{value}\\
	&\quad=\sup_{\tau \in [0, 1,\cdots,T-t]} E^Q_t\left[ e^{-r\tau }\left( \sum_{u=0}^{\tau-1}\frac{S_{t+\tau}}{S_{t+u}}cL_{t+u} + w\frac{S_{t+\tau}}{S_{t}}-b(t+\tau)\ddot{a}(T)L_{\tau+t-1}e^{-r(T-\tau-t)}\right)^+ \,\right]
	\nonumber
	\end{align}

	At time $t$, given that the DC account balance is $w$, we define the \emph{exercise value} of the option, denoted $v_1^e(t,w)$, as the value if the member decided to switch at that date, and the \emph{continuation value}, denoted $v_1^h(t,w)$ as the value if the member decides not to switch. Then
	\begin{align}
	v_1^e(t, w) &=  \left(w - tbL_{t-1}\ddot{a}(T)e^{-r(T-t)}\right)^+\\
	v_1^h(t,w) &= E^Q_t\left[e^{-r}v\left(t+1, (w+cL_{t})\frac{S_{t+1}}{S_{t}} \right)\right] \label{hvalue}
	\end{align}
	
	The option is not analytically tractable. In the next section, we present some of the properties of the value function, which will enable us to use numerical solution methods.

	\subsection{The value function at $T-1$}
	Since the option value at time $T-1$ depends only on the stock performance in the period $[T-1,T]$, it can be solved using the Black-Scholes Formula for the continuation value
	\begin{align*}
	v_1^h(T\!-\!1, w) &= E^Q\left[e^{-r}\left((w+cL_{T-1})\frac{S_{T}}{S_{T-1}}-b\,TL_{T-1}\ddot{a}(T) \right)^{+}\, \right]\\
	&=N(d_1)(w+cL_{T-1})-N(d_2)b\,TL_{T-1}\ddot{a}(T)e^{-r}\\
	\mbox{where~~}d_1&= \frac{1}{\sigma_S}\left[\ln\left(\frac{(w+cL_{T-1}}{bTL_{T-1} \ddot{a}(T)}\right)+\left(r+\frac{\sigma_S^2}{2}\right)\right]\\
	d_2 &= d_1-\sigma_S
	\end{align*}
	thus
	\begin{align*}
	v_1(T-1,w) &=
	\max\left(w-b(T-1)L_{T-2}\ddot{a}(T)e^{-r}, \,v_1^h(T-1,w)\right)
	\end{align*}
	
	\subsection{The Exercise Frontier}
	The following proposition gives the general convexity and monotonicity of the value function
	\begin{prop}\label{proposition_1}
		At each observation date $0\leq t\leq T$, the value function $v_1(t,w)$ is a continuous, strictly positive, non-decreasing and convex function of w.
	\end{prop}
	The proof is given in Appendix \ref{appB}.
	
	Like other Bermudan and American-type options, there exists a continuation region $\mathbf{C}$ and a stopping region (or exercising region) $\mathbf{D}$.  When the time and DC account value pair $(t,w)$ is in the continuation region, it is optimal for the member to stay in the DC plan. When $(t,w)$ is in the stopping region, it is optimal for the member to switch to the DB plan. The option to switch is exercised when $(t,w)$ moves into the stopping region. Mathematically, the continuation  and stopping regions are defined as
	
	\begin{itemize}
		\item $v_1^h(t,w)>v_1^e(t,w) \Leftrightarrow (t,w)\in \mathbf{C}$
		\item $v_1^h(t,w)\leq v_1^e(t,w) \Leftrightarrow (t,w)\in \mathbf{D}$	
	\end{itemize}

	\begin{prop}\label{proposition_2}
		There exists a function $\varphi(t)$ such that
		\begin{align*}
		v_1(t,w) = \left\{\begin{matrix}
		v_1^e(t,w) & \text{if } w\geq\varphi(t)\\
		v_1^h(t,w) &\text{if } w<\varphi(t)
		\end{matrix}\right.
		\end{align*}
	\end{prop}
	The proof is shown in Appendix \ref{appC}. The function $\varphi(t)$ is the exercise boundary that separates continuation and exercise regions. Notice, it is possible, under certain parameters, that $\varphi(t) = \infty$ for some $t<T$, which means that it is not optimal to switch regardless of the DC account value. The next proposition will specify the situations in which $\varphi(t)< \infty$.
	\begin{prop}\label{proposition_3}
		The behavior of the exercise boundary $\varphi(t)$ depends on the ratio
$\displaystyle{ \frac{c}{b\ddot{a}(T)e^{-rT}}}$ as follows.
		\renewcommand{\theenumi}{(\roman{enumi})}
		\begin{enumerate}
			\item If $\displaystyle{\frac{c}{b\ddot{a}(T)e^{-rT}}< 1}$, then $\varphi(t)<\infty, \forall t\in [0,T]$
			\item If $\displaystyle{\frac{c}{b\ddot{a}(T)e^{-rT}}\geq 1}$, there exists a $t_{*}\in[0,1,\cdots,T-1]$, such that $\varphi(t) = \infty, \forall t\leq t_{*}$, and $\varphi(t)<\infty, \forall t>t_{*}$.
\item  The value of $t_{*}$ is  $\lfloor{t'}\rfloor$, where $t'$ satisfies
			\begin{align*}
			(t'+1)bL_{t'}\ddot{a}(T)e^{-r(T-t')} - t'bL_{t'-1}\ddot{a}(T)e^{-r(T-t')}-cL_{t'} = 0
			\end{align*}
				\item If $\displaystyle{c>b\ddot{a}(T) \left(\left(1-e^{-\mu_L}\right)T+e^{-\mu_L}\right)e^{-r}}$ then $t_{*} = T-1$, which means $\varphi(t) = \infty$, $\forall t<T$
				\item If $\displaystyle{\frac{c}{b\ddot{a}(T)e^{-rT}} =  1}$ then $t_{*} = 0$.
			\end{enumerate}

	\end{prop}
	The proof is given in the Appendix \ref{appD}. This proposition  demonstrates that the ratio between the DC contribution rate and the accrual rate of the DB benefit will determine the shape of the exercise boundary. In the extreme case, when the DC contribution rate is much higher than the DB accrual rate, it is optimal for the employee to wait until the retirement date, and the Bermudan-DB underpin option simplifies to a DB underpin plan.
	
	\section{Bermudan DB underpin plan - Continuous Case}
	By extending the current model to incorporate a stochastic salary process, solving the value of the option becomes a three dimensional problem. The optimal exercise boundary will depend on both the salary and the DC account balance. In this section, we reconstruct the problem in a  continuous setting, with a stochastic salary.
	
	The mathematical formulation  replaces the summation sign with the integral sign, and extends the admissible stopping time set to $[t,T]$. By the Optional Sampling Theorem, we can define the value function similarly to the discrete case.

	\begin{align*}
	v_2(t,w,l) &=\sup_{0\leq \tau \leq T-t} E^Q_t\bigg[e^{-r\tau }\left(W_{t+\tau}-K_{t+\tau}\right)^+ \bigg|W_t=w, L_t =l \bigg]\\
	\text{where}\\
	dW_t&= (rW_t + cL_t)dt + \sigma_S W_tdZ^Q_S(t)\\
	K_t &= b\,t\,\ddot{a}(T)\,L_t\,e^{-r(T-t)}
	\end{align*}
	Notice now that the value function also depends on the salary at time t. In the following sections we show that although the value function is three dimensional, the optimal exercise boundary is still only two dimensional, in the sense that the value function only depends on the wealth-salary ratio process.

	\subsection{Definition of the Value Function}
	We  define a new variable $Y_t = \frac{W_t}{L_t}$, which represents the DC wealth-salary ratio. Using Girsanov's theorem, we can show that the value function ${v_2(t,w,l) = l\, v_2(t,w/l)}$, where $v_2(t,y)$ is defined as
	\begin{align*}
	v_2(t,y) = \sup_{0\leq \tau \leq T-t}E^{\tilde{P}}\left[\left(Y_{t+\tau}^{t,y}-b(t+\tau)\ddot{a}(T)e^{-r(T-t-\tau)}\right)^{+}|{\mathcal{F}}_t \right]
	\end{align*}
	and the wealth-salary ratio follows
	\begin{align*}
	dY_t = cdt + \sigma_YY_tdZ_Y^{\tilde{P}}(t)
	\end{align*}
	where $Z_Y^{\tilde{P}}(t)$ is standard Brownian Motion under the measure $\tilde{P}$. Details of the derivation can be found in  Appendix \ref{appE}.
	
	\subsection{Properties of the Value Function and Exercise Frontier}
	\begin{prop}\label{proposition_4}
	The value function $v_2(t,y)$ is a continuous, strictly positive, non-decreasing and convex function of y, and a continuous and non-increasing function of time t.
	\end{prop}
	The proof can be found in Appendix \ref{appF}.
	
	As above, we denote the continuation region $\mathbf{C} = \left\{(t,y)\in (0,T)\times [0,\infty): v_2(t,y)>v^e_2(t,y) \right\}$ and stopping region $\mathbf{C} = \left\{(t,y)\in (0,T)\times [0,\infty): v_2(t,y)=v^e_2(t,y) \right\}$, and let $\varphi(t)$ be the exercise frontier which separates the two regions.

	\begin{prop}\label{proposition_5}
		There exists a function $\varphi(t)$ such that
		\begin{align*}
		v_2(t,y) = \left\{\begin{matrix}
		v_2^e(t,y) & \text{if } y\geq\varphi(t)\\
		v_2^h(t,y) &\text{if } y<\varphi(t)
		\end{matrix}\right.
		\end{align*}
	\end{prop}
	The proof can be found in Appendix \ref{appF}.
	
\begin{figure}[H]
		\centering
		\includegraphics[width=6.5in]{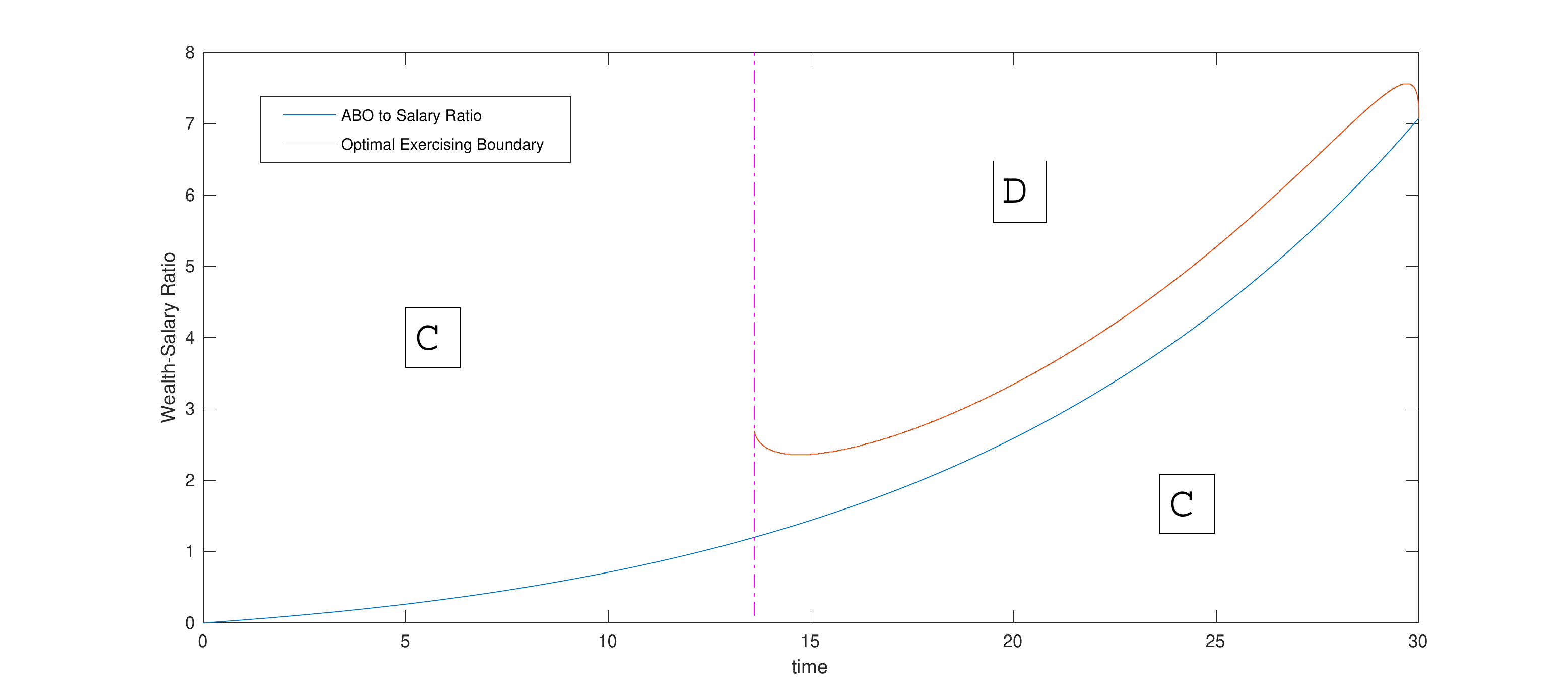}
		\caption{Example of Optimal Exercising Boundary, $r = 0.06$, $\sigma_S = 0.15$, $\sigma_L = 0.04$, $c=0.16$, $b = 0.016$ and $a_T = 14.75$}
		\label{fig_boundary_example}
	\end{figure}

	For numerical solution, we formulate the problem as a free-boundary PDE problem.
	\begin{align*}
	&\frac{\partial v_2}{\partial t} + c\frac{\partial v_2}{\partial y}+\frac{\sigma_Y^2 y^2}{2}\frac{ \partial^2v_2}{\partial y^2} = 0 \quad in \quad \mathbf{C}\\
	&v_2(t,y) = v_2^e(t,y) \quad in \quad \mathbf{D}\\
	&v_2(t,y) > v_2^e(t,y) \quad in \quad \mathbf{C}\\
	&v_2(t,y) = v_2^e(t,y) \quad for \quad y = \varphi(t) \quad or \quad t = T
	\end{align*}
	In appendix \ref{appF} we illustrate the condition where $\varphi(t) = \infty$, and we prove that when the DC contribution rate is very high, such that ${c>b\ddot{a}_T(1+rT)}$, then the option becomes the regular DB underpin option. Figure \ref{fig_boundary_example} shows an example of  the optimal exercise boundary.
	
	\subsection{Different Discount Rates for ABO Calculation}
	In practice, actuaries often use a discount rate higher than the observed market risk-free rate to determine the ABO. Here we denote the $\gamma$ as the discount rate for the ABO calculation, so the ABO has the following form:
	\begin{align*}
	K_t = tbL_t\ddot{a}(T)e^{-\gamma (T-t)}
	\end{align*}
	It is not difficult to show that the new cost function can be expressed in the same form as equation (\ref{cost_function}) through the Optional Sampling Theorem. Under our stochastic salary assumption, however, the risk free rate $r$ only appears in the ABO calculations, which means that $\gamma$ and $r$ are mathematically indistinguishable. Thus, we only include $\gamma$ in our study of deterministic salary assumptions. The new value function preserves similar properties (with slight modifications) as those stated in Propositions (\ref{proposition_1}), (\ref{proposition_2}), (\ref{proposition_3}) for the discrete case, and (\ref{proposition_4}) and (\ref{proposition_5}) in the continuous case. Notice that using different discount rates for the ABO does not violate our market-consistent valuation principle. Here we use $v_4$ to denote the price with $\gamma$ as discount rate for ABO:
	\begin{align*}
	v_4(t,w) = \sup_{0\leq \tau \leq T-t}E^{Q}\left[e^{-r\tau}\left(W_{t+\tau}^{t,w}-b(t+\tau)L_{t+\tau}\ddot{a}(T)e^{-\gamma(T-t-\tau)}\right)^{+}|{\mathcal{F}}_t \right]
	\end{align*}

	\section{Numerical Examples}
	In this section we present numerical results for the values of the Bermudan DB Underpin plan. For the continuous setting, we use the penalty method (see \citet{Forsyth02quadraticconvergence}) and for the discrete case, we use the Least Square Method (see \citet{Longstaff01valuingamerican}). We compare the Bermudan option with the DB underpin and with the Florida Second Election option. For the DB underpin option, to be consistent with the valuation method for Bermudan DB underpin option, we evaluate it using Monte Carlo simulation in the discrete setting and the Crank Nicolson finite difference method in the continuous setting. For the second election option, we are able to derive explicit solutions under both discrete and continuous settings (see Appendix \ref{appG}). Furthermore, we include the Bermudan option with deterministic salary in the continuous setting (denoted by $v_3$) as an intermediate comparison with the discrete case.
	
	\subsection{Benchmark Scenario}
	The initial benchmark parameter set is as follows. Later we perform some sensitivity tests on each of the parameters.

	\begin{itemize}
		\item $\mu_L = r = 0.04$. We set $\mu_L$ to be the same as the risk free rate, to make consistent comparison with the stochastic salary assumption (when salary is assumed to be hedgeable).
		\item $\rho = 0$, assumes no correlation between salary and investment return.
		\item $\sigma_S = 0.15$,  $b = 0.016$, $c = 0.125$ and $ \ddot{a}(T) = 14.75$.
		\item $L_0 = 1$, $t= 0$, and $W_0 = 0$, so that all values are given per unit of starting salary.
	\end{itemize}
	
	\subsection{Cost}
	Table \ref{table_cost_continuous} shows the value of each pension plan in the continuous setting. The price for the Bermudan DB-Underpin (BDBU), the Florida Second Election (FSE) and the DB underpin plans are expressed as an additional cost on top of the base DB plan. We show values for the Bermudan DB with stochastic salaries $v_2$, and with deterministic salaries, $v_3$.  Table \ref{table_cost_discrete} shows the results in the discrete setting.\\
	
	\begin{table}
		\centering
		\begin{tabular}{|c|c|c|c|c|c|c|}
			\hline
			Time to Retirement & DB & DC  &FSE & DB-Underpin &BDBU &BDBU \\
			$\tau$ & & & & &$v_2$ & $v_3$\\\hline
			10yr &2.36	&1.25		&0		&0.0023	&0.0070&0.0062\\
			15yr &3.54	&1.87		&0		&0.0126	&0.0354&0.0315\\
			20yr &4.72	&2.50		&0.0203	&0.0348	&0.1010&0.0936\\
			30yr &7.08	&3.75	&0.2179	&0.1199	&0.3492 & 0.3355\\
			40yr &9.44	&5.00		&0.5837	&0.2594	& 0.7380 &0.7194\\
			\hline
		\end{tabular}	
		\caption{Cost of each pension plan per unit of starting salary, continuous setting. FSE is the Florida second election option; and BDBU $v_2$ is the Bermudan DB underpin with stochastic salaries, and BDBU $v_3$ is the Bermudan DB underpin with deterministic salaries.  Hybrid costs are additional to the basic DB cost.}
		\label{table_cost_continuous}
	\end{table}
\begin{table}
		\centering
		\begin{tabular}{|c|c|c|c|c|c|}
			\hline
			Time to Retirement & DB & DC & FSE & DB-Underpin & BDBU \\
			$\tau$ & & & & &$v_1$ \\\hline	\hline
			10yr    &2.2675    &1.2500      &      0  &  0.0039   (0.0011)& 0.0099  (0.0001)\\
			15yr    &3.4012    &1.8750      &      0  &  0.0210   (0.0020)& 0.0456  (0.0003)\\
			20yr    &4.5349    &2.5000      & 0.0304  &  0.0458   (0.0029)& 0.1190  (0.0006)\\
			30yr    &6.8024    &3.7500      & 0.2476  &  0.1455   (0.0048)& 0.3752  (0.0014)\\
			40yr    &9.0699    &5.0000      & 0.6280  &  0.3115   (0.0069)& 0.7726  (0.0025)\\
			\hline
		\end{tabular}	
		\caption{Cost of each pension plan, discrete setting. FSE is the Florida second election option, and BDBU is the Bermudan DB underpin. Hybrid costs are additional to the basic DB cost.}
		\label{table_cost_discrete}
	\end{table}

	Some observations can be made:
	\begin{itemize}
		\item BDBU $v_2$ is greater than BDBU $v_3$ which reflects the additional costs from stochastic salaries, but the results are fairly close. We would expect that in the discrete case, a stochastic salary would lead to a higher but close cost as in the deterministic salary assumption.
		\item When the expected retirement date is near, under the benchmark parameters, participants in the Second Election plan should always be in the DB plan ($t^{*} = 0$), so there will be no extra cost required to fund  the second election option.
		\item Although the Bermudan DB underpin option is greater than both the DB underpin and the Second Election, none of the values in the benchmark scenario exceed 10\% of the cost of a DB plan. For horizons of 30 years or less, the extra cost from the DBDU option costs around 5\% more than the basic DB plan.
		\item As $T$ increases, the cost is increasing, at an increasing rate, for all three options. The underpin and election options become significantly more costly over 40 years compared with the cost for 30 years.
\item In the discrete case, the value of the options are generally greater than in the continuous case, and we would expect a larger difference if stochastic salary is incorporated. However, the observations made in the continuous case also apply in the discrete case.
\end{itemize}	
	\subsection{Sensitivity Tests}
	In this section, we present the sensitivity tests over all parameters in the continuous setting. We consider $c$, $\mu_L$, $r$, $\sigma_S$, $\gamma$ and $b$ under the deterministic salary assumption and $\sigma_L$ and $\rho$ under the stochastic salary assumption. We also fix the time horizon to be 30 years. Details are displayed in Table \ref{sensitivity_deterministic}.
	
	\begin{table}	
		\centering
		\begin{tabular}{|c|c|c|c|c|c|c|c|c|c|}
			\hline
			Factor & \multicolumn{9}{|c|}{Sensitivity Tests}\\
			\hline
		$r$		&0			&0.01	&0.02		&0.03	&\textbf{0.04}	&0.05	&0.06	&0.07 	&0.08\\
			\hline
		DB		&23.5064	&17.414	&12.9006	&9.557	&7.08	&5.245	&3.8856	&2.8785	&2.1325\\
	$v^{se}$	&0			&0		&0			&0.598	&0.2179	&0.4045	&0.5786	&0.7213	&0.8276\\
		$v_3$	&0.0174		&0.0437	&0.1018		&0.2023	&0.3355	&0.451	&0.6202	&0.7398	&0.8327\\
		$v_U$	&0.0093		&0.0198	&0.039		&0.0711	&0.1199	&0.1878	&0.2741	&0.374	&0.4797	\\
			\hline
		$\mu_L$	&0		&0.01	&0.02	&0.03	&\textbf{0.04}	&0.05	&0.06		&0.07	&0.08\\
			\hline
		DB		&2.1325	&2.8785	&3.8856	&5.245	&7.08	&9.557	&12.9006	&17.414	&23.5064\\
	$v^{se}$	&0.3448	&0.2987	&0.265	&0.2389	&0.2179	&0.2005	&0.1858		&0.1732	&0.1623\\
		$v_3$	&0.5299	&0.473	&0.4217	&0.3758	&0.3355	&0.3004	&0.2703		&0.2446	&0.2229\\
		$v_U$	&0.4797	&0.374	&0.2741	&0.1878	&0.1199	&0.0711	&0.039		&0.0198	&0.0093	\\
			\hline
		$c$		&0.085	&0.095	&0.105	&0.115	&\textbf{0.125}	&0.135	&0.145	&0.155	&0.165\\
			\hline
	$v^{se}$	&0.0163	&0.0466	&0.0909	&0.1484	&0.2179	&0.2987	&0.3902	&0.4917	&0.6026\\
		$v_3$	&0.0759	&0.1235	&0.1831	&0.2539	&0.3355	&0.4271	&0.5282	&0.6383	&0.757\\
		$v_U$	&0.0183	&0.0325	&0.0533	&0.0821	&0.1199	&0.1679	&0.2269	&0.2975	&0.3801\\
			\hline	
	$\sigma_S$	&0.07	&0.09	&0.11	&0.13	&\textbf{0.15}	&0.17	&0.19	&0.21	&0.23\\
			\hline
		$v_3$	&0.2205	&0.2314	&0.2542	&0.2893	&0.3355	&0.391	&0.4538	&0.5225	&0.5954\\
		$v_U$	&0.0012	&0.0094	&0.0311	&0.0685	&0.1199	&0.183	&0.2552	&0.3341	&0.418\\
			\hline
		$b$		&0.012	&0.013	&0.014	&0.015	&\textbf{0.016}	&0.017	&0.018	&0.019	&0.02\\
			\hline
		DB		&5.31	&5.7525	&6.195	&6.6375	&7.08	&7.5225	&7.965	&8.4075	&8.85\\
	$v^{se}$	&0.4665	&0.3896	&0.3235	&0.2667	&0.2179	&0.17	&0.1401	&0.1096	&0.0838\\
		$v_3$	&0.5826	&0.5075	&0.4422	&0.3853	&0.3355	&0.2918	&0.2535	&0.2199	&0.1904\\
		$v_U$	&0.2958	&0.2342	&0.1864	&0.1491	&0.1199	&0.0969	&0.0788	&0.0643	&0.0527\\
			\hline
		$\gamma$	&0		&0.01	&0.02	&0.03	&\textbf{0.04}	&0.05	&0.06	&0.07	&0.08\\
			\hline
	$v^{se}$	&0		&0		&0		&0.067	&0.2179	&0.3792	&0.5281	&0.6591	&0.7728\\
		$v_3$	&0.1315	&0.1492	&0.1835	&0.2442	&0.3355	&0.4513	&0.5811	&0.715	&0.8463\\
			\hline
		\end{tabular}
		\caption{Sensitivity test over ($c$, $\mu_L$, $r$, $\sigma_S$, $\gamma$ and $b$), deterministic salary assumption; $v^{se}$ is the additional FSE cost, $v_3$ is the additional BDBU cost, and $v_U$ is the additional DB underpin cost.}
		\label{sensitivity_deterministic}
	\end{table}
	
	We note the following points.
	\begin{itemize}
		\item All three plans share the same directional trends as the parameters change, but with some very different sensitivities.
\item The additional costs of the hybrid plans  react in the opposite direction to the underlying DB costs, for parameters which influence the DB cost. For example, increasing the risk free rate $r$ decreases the DB cost, but increases the additional hybrid costs. Increasing the accrual rate $b$ increases the DB costs, but decreases the additional hybrid costs. Hence, the sensitivity of the total costs to the changing parameters is rather more muted than the sensitivity of the additional costs shown in the table.
		\item The risk-free rate has a significant impact on the Bermudan DB underpin option value, especially on the relative cost with respect to the DB plan. However, as shown by the sensitivity test on $\gamma$, the impact mostly comes from the value of the ABO. Also, it is interesting to note that when $r$ is high, the cost of the FSE option is quite close to the Bermudan DB underpin option.
		\item Decreases in the accrual rate $b$ for the DB plan increase the relative value of the Bermudan DB  option. The extra cost reflects the fact that the funding of the DC plan is higher than the DB plan. When $b$ is high, the fast accumulation of the DB benefit would discourage employees from entering into the DC account, and the option value will be reduced.
		\item In our model, the cost of the second election option is independent of the market volatility $\sigma_S$.
	\end{itemize}
	Although the structure of Bermudan DB underpin plan provides more flexibility and protection to the employee than both the second election option and the DB underpin plan,  the additional cost does not appear as large as one may expect. In most scenarios, the cost is less than 5\% of the DB plan. The relative cost is high when the risk-free rate increases, however, it is interesting to note that the cost is very close to the second election option. Similarly, when the salary growth rate is low, the cost is quite close to that of the DB underpin option. This demonstrates that a very large cost comes either from the option which allows employees to switch to DC, or from the guarantee.
	
	Table \ref{sensitivity_stochastic} displays the sensitivity test for the stochastic salary parameters.
		\begin{table}	
			\centering
			\begin{tabular}{|c|c|c|c|c|c|c|c|c|c|}
				\hline
				Value & \multicolumn{9}{|c|}{Sensitivity Tests}\\
				\hline
		$\sigma_L$	&0.01	&0.02	&0.03	&0.04	&0.05	&0.06	&0.07	&0.08	&0.09\\
				\hline
			$v_3$	&0.3363	&0.3389	&0.3432	&0.3492	&0.357	&0.3665	&0.3778	&0.391	&0.4058\\
			$v_{U}$	&0.1209	&0.1238	&0.1286	&0.1354	&0.1443	&0.1551	&0.1681	&0.183	&0.2001\\
				\hline
			$\rho$	&-1		&-0.9	&-0.5	&-0.1	&0		&0.1	&0.5	&0.9	&1\\
				\hline
			$v_3$	&0.4538	&0.4434	&0.4015	&0.3596	&0.3492	&0.3389	&0.298	&0.2623	&0.2542\\
			$v_{U}$	&0.2552	&0.2432	&0.195	&0.1472	&0.1354	&0.1238	&0.079	&0.0396	&0.0311\\
				\hline		
			\end{tabular}
			\caption{Sensitivity Test over ($\sigma_L$ and $\rho$), stochastic salary}
			\label{sensitivity_stochastic}
		\end{table}
	The risks involved in the stochastic salary process have less impact overall. Larger volatility in the salary process, as well as negative correlation between the salary and equity market, will increase the volatility of wealth-salary ratio process, and thus, increases the option value. These two risks have same effect on the DB underpin option, which has been previously observed by \citet{chen2009}. However, under our assumptions, the value of second election option is immunized to these two risks.
	
	\section{Conclusion and Future Work}
	In this paper, we discuss a new pension design, which combines Florida's second election option and the DB underpin option, to form a Bermudan-type DB underpin plan. We summarize some key characteristics of the option, such as convexity and monotonicity. Also, we provide illustrations of the behavior of the early exercise region, and specifically include the situation where the Bermudan DB underpin simplifies to the DB underpin plan.
	
	Our numerical illustrations demonstrate that, although the Bermudan-type DB underpin option may end up costing more than both the DB underpin option and the second election option combined, it does not cost more than 10\% of the DB plan in general. In cases when the relative cost of the option compared with the DB plan is very large, for example when risk-free rate is high or salary growth rate is low, the actual cost of the Bermudan DB underpin plan is indeed smaller.
	
	The Bermudan DB underpin plan shifts more risk and cost to the employer  compared with a DB plan, so it may not be an attractive  option for pension sponsors, but the overall costs can be managed to some extent by varying the DC contribution rate and the DB accrual rate. Furthermore, it offers an attractive portable benefit for younger employees, which should help with attracting new staff, and offers a substantial retention benefit for older employees, which should help with reducing turnover of midcareer staff. It provides predictable income in retirement.  In addition, we have used the Bermudan DB underpin to connect the FSE plan design to the DB underpin.
	
There are many outstanding questions that we hope to address in future work.
	\begin{itemize}
		\item The assumption of a complete market may be too strong. It may also be interesting to consider the situation when salary can only be partially hedged.
		\item The sensitivity of the option value of the risk-free rate indicates that the results might be sensitive to stochastic interest rates.
		\item The annuity factor is highly sensitive to the choice of discount rate. However, as we assume a constant annuity factor, we may underestimate the risk.
		\item It may also be interesting to explore the option from the employee's perspective. For example, evaluating the utility gains from the guarantee.
	\end{itemize}
	
	\section{Acknowledgments}
	We are very grateful to Professor Chen Xinfu, particularly for a thoughtful discussion on the structure of the pension plan and the PDE formulation.
		
\bibliography{Bib_Bermudan_DB_Underpin}
\bibliographystyle{jss}

	\appendix
\section{Appendix}
	\subsection{Cost Function - Discrete Case}
	Here is the derivation of equation (\ref{cost_function}):
	\begin{align*}
&	C(t,w) =\sup_{\tau \in [0,1,\cdots, T-t]} E^Q\Bigg[\sum_{u=0}^{\tau-1}e^{-ru}cL_{t+u} +  e^{-r\tau}\left(K_Te^{-r(T-\tau-t)}-K_{t+\tau}e^{-r(T-t-\tau)}\right) \\
	& \qquad \qquad \qquad  +e^{-\tau r}\left(K_{t+\tau}e^{-r(T-\tau-t)} - W_{\tau+t}^{t,w}\right)^+  \bigg|{\mathcal{F}}_t\Bigg]\\
	=&\sup_{\tau \in [0,1,\cdots, T-t]} E^Q\left[\sum_{u=0}^{\tau-1}e^{-ru}cL_{t+u} +  e^{-r\tau}\left(K_Te^{-r(T-\tau-t)}-K_{t+\tau}e^{-r(T-t-\tau)}\right) \right.\\
	& \qquad \qquad \qquad \left. +K_{t+\tau}e^{-r(T-t)} - e^{-r\tau }W_{\tau+t}^{t,w} + e^{-r\tau }\left(W_{\tau+t}^{t,w}-K_{t+\tau}e^{-r(T-t-\tau)}\right)^+  \bigg|{\mathcal{F}}_t\right]\\
	=&\sup_{\tau \in [0,1,\cdots, T-t]} E^Q\left[\sum_{u=0}^{\tau-1}e^{-ru}cL_{t+u} +  K_Te^{-r(T-t)}  - e^{-r\tau }W_{\tau+t}^{t,w} + e^{-r\tau }\left(W_{\tau}^{t,w}-K_{t+\tau}e^{-r(T-t-\tau)}\right)^+ \bigg|{\mathcal{F}}_t \right]\\
	\end{align*}
	To further reduce our equation, we need the Optional Sampling Theorem. First, observe that
	\begin{align*}
	E^Q\left[e^{-r(t-s)}W_{t}^{s,w} |{\mathcal{F}}_s\right] &= E^Q\left[e^{-r(t-s)}\left(w\frac{S_t}{S_s} + \sum_{u=s}^{t-1}\frac{S_t}{S_u}cL_u\right) \bigg|{\mathcal{F}}_s\right]\\
	& = w+ e^{-r(t-s)} \sum_{u=s}^{t-1}e^{r(t-u)}cL_u=w+\sum_{u=s}^{t-1}e^{r(s-u)}cL_u
	\end{align*}
	Define a new process $X_t$ as
	\begin{align*}
	X_t &= e^{-rt}W_t^{0,w} - \left(w + \sum_{u=0}^{t-1}e^{-ru}cL_u\right)\\
	&=  e^{-tr}w\frac{S_t}{S_0}- w + \sum_{u=0}^{t-1}\left(e^{-rt}\frac{S_t}{S_u}-e^{-ru}\right)cL_u\\
	\end{align*}
	It is easy to verify that $X_t$ is a martingale:
	\begin{align*}
	E^Q\left[X_t|{\mathcal{F}}_s\right] &=e^{-rt}w\frac{S_s}{S_0}e^{r(t-s)} -w + \sum_{u=0}^{s}\left(e^{-rt}\frac{S_s}{S_u}e^{r(t-s)} - e^{-ru}\right)cL_u  +\sum_{u=s+1}^{t-1}\left(e^{-rt}e^{r(t-u)}-e^{-ru}\right)cL_u \\
	&=e^{-rs}w\frac{S_s}{S_0} - w +\sum_{u=0}^{s-1}\left(e^{-rs}\frac{S_s}{S_u}-e^{-ru}\right)cL_u \\
	&=X_s
	\end{align*}
	Let $\tau \in [0,1,\cdots,T-t]$ be any stopping time, by optional sampling theorem, we have
	\begin{align*}
	 &E^Q\left[X_{\tau+t} |{\mathcal{F}}_t\right]= X_t\\
	&\implies E^Q\left[\underbrace{e^{-r(\tau+t)}w\frac{S_{t+\tau}}{S_0} + \sum_{u=0}^{t+\tau-1}e^{-r(t+\tau)}\frac{S_{t+\tau}}{S_u}cL_u|{\mathcal{F}}_t}_{=e^{-r(\tau+t)}W_{t+\tau}^{t,W_t}} \right]= E^Q\left[w+\sum_{u=0}^{t+\tau-1}e^{-ru}cL_{u}|{\mathcal{F}}_t\right] + X_t\\
&	E^Q\left[e^{-r\tau}W_{t+\tau}^{t,W_t}|{\mathcal{F}}_t\right]= E^Q\left[\sum_{u=0}^{\tau-1}e^{-ru}cL_{u+t}|{\mathcal{F}}_t\right] + W_t
	\end{align*}
	Substitute the last line into cost function, we have
	\begin{align*}
	C(t,w) =&\sup_{\tau \in [0,1,\cdots, T-t]} \left\{E^Q\left[\sum_{u=0}^{\tau-1}e^{-ru}cL_{t+u}\right] - E^Q\left[e^{-r\tau}W_{\tau+t}^{t,w}|{\mathcal{F}}_t \right]+ E^Q\left[  K_Te^{-r(T-t)} \right] \right.\\
	& \qquad \qquad \left. +E^Q\left[ e^{-r\tau}\left(W_{\tau}-K_{t+\tau}e^{-r(T-t-\tau)}\right)^+ \right]\right\}\\
	&=\underbrace{ E^Q\left[K_Te^{-r(T-t)}\right]}_{\text{The ABO of DB plan at t}} + \underbrace{\sup_{\tau \in [0,1,\cdots, T-t]}E^Q\left[ e^{-r\tau}\left(W_{\tau+t}^{t,w}-K_{t+\tau}e^{-r(T-t-\tau)}\right)^+|{\mathcal{F}}_t \right]}_{\text{Price of the Option}} -w
	\end{align*}
	
	\subsection{Characteristics of Value Function - Discrete Case\label{appB}}
	Here is the proof of Proposition (\ref{proposition_1})

	\subsubsection{Value function is non-decreasing in w\label{appB1}}
	For $h>0$, we have $(x-k)^+ - (x+h-k)^+\leq 0$, for all x, therefore,
	\begin{align*}
	v(t,w)& - v(t,w+h) \\
&\!\!\leq \sup_{\tau \in [0,1,\cdots, T-t]} E^Q\left[e^{-r\tau}\left\{\left(W_{t+\tau}^{t,w}-K_{t+\tau}e^{-r(T-t-\tau)}\right)^+-\left(W_{t+\tau}^{t,w} + h\frac{S_{t+\tau}}{S_t}-K_{t+\tau}e^{-r(T-t-\tau)}\right)^+\right\}\right]\\
	&\leq 0
	\end{align*}
	since $\frac{S_{t+\tau}}{S_t}$ is strictly positive a.s.. Notice, although the value function is increasing in the initial DC balance, the cost function $C(t,w)$ is the opposite.
	
	\begin{align*}
	&C(t,w+h) - C(t,w) \\
	&\leq \sup_{\tau \in [0,1,\cdots, T-t]} E^Q\left[e^{-r\tau}\left\{\left(W_{t+\tau}^{t,w}+ h\frac{S_{t+\tau}}{S_t}-K_{t+\tau}e^{-r(T-t-\tau)}\right)^+-\left(W_{t+\tau}^{t,w} -K_{t+\tau}e^{-r(T-t-\tau)}\right)^+\right\}\right]-h\\
	&\leq \sup_{\tau \in [0,1,\cdots, T-t]} E^Q\left[e^{-r\tau}\left\{h\frac{S_{t+\tau}}{S_t}+\left(W_{t+\tau}^{t,w}-K_{t+\tau}e^{-r(T-t-\tau)}\right)^+-\left(W_{t+\tau}^{t,w} -K_{t+\tau}e^{-r(T-t-\tau)}\right)^+\right\}\right]-h\\
	& = 0
	\end{align*}

	\subsubsection{$v^h(t,w)$ and $v^e(t,w)$ are non-decreasing in w\label{appB2}}
	For $x>y$,
	\begin{align*}
	v^h(t,x)-v^h(t,y) &= e^{-r}E^Q\left[v\left(t+1, (x+cL_t)\frac{S_{t+1}}{S_t}\right)-v\left(t+1, (y+cL_t)\frac{S_{t+1}}{S_t}\right)\right]\\
	&\geq 0
	\end{align*}
	Thus, $v^h$ is increasing in the DC account balance. It is also easy to see that
	\begin{align*}
	v^e(t,x) - v^e(t,y) &= (x-K_te^{-r(T-t)})^+ -(y-K_te^{-r(T-t)})^+\\
	&\geq 0
	\end{align*}
	
	\subsubsection{Continuity of value function in $ w$\label{appB3}}
	For $x>y$, using the fact that $\sup[X] - \sup[Y] \leq \sup[X-Y]$ and $(x-k)^+ - (y-k)^+ \leq (x-y)^+$, we have
	\begin{align*}
	&| v(t,x) - v(t,y)| \\
	&\leq\Bigg|\sup_{\tau \in [0,1,\cdots, T-t]}E^Q\left[ e^{-\tau r} \left(x\frac{S_{t+\tau}}{S_{t}} + \sum_{u=0}^{\tau}\frac{cS_{t+\tau}L_{u+t}}{S_{u+t}} - K_{t+\tau}e^{-r(T-\tau-t)}\right)^+ \right.\\
	&\qquad \qquad \qquad\left. -e^{-\tau r} \left(y\frac{S_{t+\tau}}{S_{t}} + \sum_{u=0}^{\tau}\frac{cS_{t+\tau}L_{u+t}}{S_{u+t}}- K_{t+\tau}e^{-r(T-\tau-t)}\right)^+
	\big|{\mathcal{F}}_t\big.\right]\Bigg|\\
	& \qquad \qquad \qquad \\
	&\leq  \sup_{\tau \in [0,1,\cdots, T-t]}E^Q\left[ e^{-r\tau}\left((x-y)\frac{S_{t+\tau}}{S_{t}} \right)  \right] \\
	&\leq (x-y)
	\end{align*}
	Thus, $v$ is Lipschitz continuous in $w$ and clearly for any $\epsilon>0$
	\begin{align}\label{formula_increment}
	v(t,w+\epsilon)\leq  v(t,w)+\epsilon
	\end{align}
	The similar property for $v^h$ follows immediately, for any $\epsilon>0$
	\begin{align*}
	v^h(t,w+\epsilon) &= E^Q\left[e^{-r}v\left(t+1,(w+\epsilon+cL_t)\frac{S_{t+1}}{S_t}\right)\right]\\
	&\leq  E^Q\left[e^{-r}v\left(t+1,(w+cL_t)\frac{S_{t+1}}{S_t}\right) + e^{-r}\epsilon\frac{S_{t+1}}{S_{t}}\right]\\
	&= v^h(t,w) + \epsilon
	\end{align*}
	
	\subsubsection{Convexity of the value function\label{appB4}}
	We follow \cite{Hatem2002}, and prove the convexity by induction. For any $w_1>0$ and $w_2>0$, and $0\leq \lambda\leq 1$
	\begin{align*}
	v^{h}(T-1,\lambda w_1 + (1-\lambda)w_2) &= E^Q\left[e^{-r}\left((\lambda w_1 + (1-\lambda)w_2+cL_{T-1})\frac{S_T}{S_{T-1}} - bL_{T-1}T \ddot{a}(T)\right)^+\right]\\
	&\leq \lambda v^h(T-1, w_1) + (1- \lambda)v^h(T-1,w_2)
	\end{align*}
	Thus, $v^h$ is convex at time T-1, and similarly $v^e$ is convex at time $T-1$. Since
	\begin{align*}
	v(T-1,w) = \max\left(v^e(T-1,w), v^h(T-1,w)\right)
	\end{align*}
	$v$ is also convex at time $T-1$.\\
	
	We now assume that result holds for time $t+1$, where $0\leq t\leq T-2$, then the continuation value at time $ t$ is
	\begin{align*}
	v^h&(t,\lambda w_1 + (1-\lambda)w_2) \\
	&= E^Q\left[e^{-r}v\left(t+1, (\lambda w_1 + (1-\lambda)w_2+cL_{t})\frac{S_{t+1}}{S_{t}} \right)\right]\\
	&\leq E^Q\left[e^{-r}\left(\lambda v\left(t+1, ( w_1 + cL_{t})\frac{S_{t+1}}{S_{t}} \right)+(1-\lambda)v\left(t+1, ( w_2 + cL_{t})\frac{S_{t+1}}{S_{t}} \right)\right)\right]\\
	&= \lambda E^Q\left[e^{-r} v\left(t+1,(w_1+cL_{t})\frac{S_{t+1}}{S_{t}}\right)\right]+(1-\lambda)E^Q\left[e^{-r} v\left(t+1,(w_2+cL_{t})\frac{S_{t+1}}{S_{t}}\right)\right]\\
	&= \lambda v^h(t,w_1) + (1-\lambda) v^h(t,w_2)
	\end{align*}
	Thus, $v^h$ is a convex function at time $ t$. Since $v^e$ holds the convexity for all $ t$, and $v(t,w) = \max\left(v^e(t,w), v^h(t,w)\right)$ holds for all $ t$, then $v(t,w)$ is convex function at time $ t$. Then, by induction, we have proved the convexity of $v(t,w)$.
	
	\subsection{Properties of $\mathbf{C}$ and $\mathbf{D}$ - Discrete Case \label{appC}}
	Consider first the case when $w<K_t$ at time $t$. We have $v^e(t,w)=(w-K_t)^+$, so
	\begin{align*}
	&w<K_t \Rightarrow v^e(t,w) =0
	\end{align*}
		
	The  value function, given in equation (\ref{value}), is the expected value of a function bounded below by zero, and which has a positive probability of being greater than zero, which means that the expected value is strictly greater than zero. The continuation function is the expected discounted value of the 1-year ahead value function (assuming the option is not exercised immediately), so it must be strictly greater than 0.
		
	So whenever $w \in [0, K_t]$, we have $v^h(t,w)> 0 = v^e(t,w)$. Thus  if $w \in [0,K_t]$, it cannot be optimal to exercise. Therefore, in order to explore the exercise frontier,  we need only consider cases when $w>K_t$.
		
	When $w>K_t$ we have $v^e(t,w) = w-K_t$ and
	\begin{align*}
	&(t,w)\in \mathbf{D} \implies v(w,t)=v^e(t,w) \implies v(w,t)=w-K_t
	\end{align*}  From Appendix \ref{appB3}, $v(t,w)+\epsilon \geq v(t,w+\epsilon)$ and $v^h(t,w)+\epsilon \geq v^h(t,w+\epsilon)$ for any $w\in \mathbf{R}$ and $\epsilon>0$.
		
	First, assume that  $w>K_t$ and that $(t,w) \in \mathbf{D}$,
	\begin{align*}
	(t,w) \in D &\implies v^e(t,w) \geq v^h(t,w)\\
	& \implies \forall \epsilon> 0 \quad v^e(t,w+\epsilon)=v^e(t,w) +\epsilon \geq v^h(t,w)+\epsilon \geq v^h(t,w+\epsilon)\\
	& \implies (t,w+h) \in\mathbf{D}
	\end{align*}
		
	Next, assume that $w>K_t$ and that $(t,w) \in \mathbf{C}$. Note that $v^h(t,w)+\epsilon \geq v^h(t,w+\epsilon)$ for all $\epsilon>0$ implies that $v^h(t,w-\epsilon) \geq v^h(t,w)-\epsilon$
	\begin{align*}
	(t,w) \in\mathbf{C} &\implies v^h(t,w)> v^e(t,w)\\
	& \implies \forall \epsilon> 0 \quad v^h(t,w-\epsilon)\geq v^h(t,w) -\epsilon > v^e(t,w)-\epsilon =v^e(t,w-\epsilon)\\
	& \implies (t,w-\epsilon) \in \mathbf{C}
	\end{align*}
	These results show  that it is optimal for the employee to switch to the DB plan only when his/her DC account balance is above a certain threshold at each possible switching time.
	
	\subsection{Properties of $\varphi(t)$ - Discrete Case\label{appD}}
	This section provides the proof of Proposition \ref{proposition_3}. First, notice that if $\varphi(t) < \infty$, then for sufficiently large $w$
	\begin{align*}
	v(t,w) = v^e(t,w) &\geq v^h(t,w) =
	E\left[e^{-r}v\left(t+1,\left(w+cL_{t}\right)\frac{S_{t+1}}{S_{t}}\right)\right]\\
	&\geq E\left[e^{-r}v^e\left(t+1, \left(w+cL_{t}\right)\frac{S_{t+1}}{S_{t}}\right)  \right]\\
	&=  \left(w+cL_t\right)N\left(d_{1,t,w}\right) - (t+1)bL_{t} \ddot{a}(T)e^{-r(T-t-1)}e^{-r}N(d_{2,t,w})
	\end{align*}
	where
	\begin{align*}
	d_{1,t,w} &= \frac{1}{\sigma_S}\left[\ln\left(\frac{w+cL_t}{(t+1)L_{t}b \ddot{a}(T)e^{-(T-t-1)r}}\right) + \left(r+\frac{\sigma_S^2}{2}\right)\right]\\
	d_{2,t,w} &= d_{1,t,w} - \sigma_S
	\end{align*}
	which is the Black-Scholes Formula, with initial stock price $w+cL_t$ and strike value\\  ${(t\!+\!1)bL_t \ddot{a}(T)e^{-r(T-t-1)}}$. Here we define
	\begin{align*}
	f(t) &=\lim_{w\rightarrow \infty} v^e(t,w) - E\left[e^{-r}v^e\left(t+1,\left(w+cL_t\right)\frac{S_{t+1}}{S_t}\right)\right]\\
	&= \lim_{w\rightarrow \infty} v^e(t,w) - \left(\left(w+cL_t\right)N\left(d_{1,t,w}\right) - (t+1)bL_{t} \ddot{a}(T)e^{-r(T-t-1)}e^{-r}N(d_{2,t,w})\right)\\
	&=\lim_{w\rightarrow \infty} w - \left(\left(w+cL_t\right)N\left(d_{1,t,w}\right) - (t+1)bL_{t} \ddot{a}(T)e^{-r(T-t)}N(d_{2,t,w})\right) - tbL_{t-1} \ddot{a}(T)e^{-r(T-t)}\\
	&=\lim_{w\rightarrow \infty} - \left((t+1)bL_{t} \ddot{a}(T)e^{-r(T-t)}N(-d_{2,t,w}) - \left(w+cL_t\right)N\left(-d_{1,t,w}\right)\right)\\
	& \qquad \qquad -cL_t + (t+1)bL_{t} \ddot{a}(T)e^{-r(T-t)}- tbL_{t-1} \ddot{a}(T)e^{-r(T-t)}\\
	&= (t+1)bL_{t} \ddot{a}(T)e^{-r(T-t)} - tbL_{t-1} \ddot{a}(T)e^{-r(T-t)} - cL_t
	\end{align*}
	 Clearly, whenever $\varphi(t)<\infty$, we have $f(t)\geq 0$.

If $\displaystyle{\frac{c}{b \ddot{a}(T)e^{-rT}}< 1}$, we prove $\varphi(t) < \infty, \forall t\in [0,T]$ by induction.

		\textbf{At time T}, $\varphi(T) = TbL_{T-1} \ddot{a}(T) < \infty$.\\
		\textbf{At time t}, assume $\varphi(t+1)<\infty$. We observe that for sufficiently large $w<\infty$, we have $ v(t+1,w) = v^e(t+1,w)$. Next, we can show
		\begin{align*}
		& \lim_{w\rightarrow \infty} v^h(t,w) - E\left[e^{-r}v^e\left(t+1,\left(w+cL_t\right)\frac{S_{t+1}}{S_t} \right)\bigg|{\mathcal{F}}_t\right]\\
		=& \lim_{w\rightarrow \infty}  E\left[e^{-r}\left(v\left(t+1,\left(w+cL_t\right)\frac{S_{t+1}}{S_t} \right)-v^e\left(t+1,\left(w+cL_t\right)\frac{S_{t+1}}{S_t} \right)\right)\bigg|{\mathcal{F}}_t\right]\\
		=& 0
		\end{align*}
		The last line is due to the fact that if $w\geq \varphi(t+1)$
		\begin{align*}
		&v\left(t+1,w \right)-v^e\left(t+1,w \right) = 0 <\varphi(t+1)<\infty
		\end{align*}
		and for $w<\varphi(t+1)$
		\begin{align*}
		v\left(t+1,w \right)-v^e\left(t+1,w \right) \leq v^e\left(t+1,\varphi(t+1) \right) \leq \varphi(t+1) < \infty
		\end{align*}
		since $v(t+1,w)$ is an increasing function of $w$ (Appendix \ref{appB1}). Thus, the difference is bounded by $\varphi(t+1)<\infty$ and we are able to apply the Dominated Convergence Theorem.
		Next,
		\begin{align*}
		\lim_{w\rightarrow \infty}v^e(t,w) - v^h(t,w) &= \lim_{w\rightarrow \infty}v^e(t,w)-E\left[e^{-r}v^e\left(\left(w+cL_t\right)\frac{S_{t+1}}{S_t} \right)\right]\\
		&= (t+1)bL_{t}\ddot{a}(T)e^{-r(T-t)}-tb \ddot{a}(T)e^{-r(T-t)}L_{t-1} -cL_{t} \\
		&>  (t+1)bL_{t}\ddot{a}(T)e^{-r(T-t)}-tb \ddot{a}(T)e^{-r(T-t)}L_{t-1} - L_{t}b \ddot{a}(T)e^{-rT}\\
		&>  (t+1)bL_{t}\ddot{a}(T)e^{-r(T-t)}-tb \ddot{a}(T)e^{-r(T-t)}L_{t} - L_{t}b\ddot{a}(T)e^{-r(T-t)}\\
		& = 0
		\end{align*}
		Which implies $\varphi(t)< \infty$ (otherwise if $\varphi(t) = \infty$, $\lim_{w\rightarrow \infty}v^e(t,w)-v^h(t,w)\leq 0$). \\
		Thus $\varphi(t)<\infty, \forall t\in[0,T]$.\\

 For $\displaystyle{\frac{c}{b \ddot{a}(T)e^{-rT}}\geq 1}$, we split the proof into three parts.

		\begin{enumerate}
			\item[(1)] If $\frac{c}{b \ddot{a}(T)e^{-rT}}> 1$, we first prove that there exists a $t_{*}$ such that $\varphi(t) = \infty, \forall t\leq t_{*}$, then prove  that $\varphi(t) < \infty, \forall t> t_{*}$ by induction from time T to $t_{*}+1$ as above.
	
We have $f(0)<0$ so that
				\begin{align*}
				\lim_{w\rightarrow \infty} v^e(0,w) - v^h(0,w) \leq f(0) < 0
				\end{align*}
				and thus $\varphi(0) = \infty$. Also, notice we can write $f(t)$ in the form
\[f(t) = e^{\mu_Lt}\left(te^{-r(T-t)}b\ddot{a}(T)(1-e^{-\mu_L}) +b\ddot{a}(T)e^{-r(T-t)}-c \right) = e^{\mu_Lt}h(t)\]

where $h(t)$ is a strictly increasing function of time t, if both $r$ and $\mu_L$ are non-negative, with at least one of them being strictly positive. We have $f(0)<0$ and \[f\left(\frac{1}{r}{\log\left(\frac{c}{b \ddot{a}(T)e^{-rT}}\right)}\right)>0\]
\[\mbox{where }\frac{1}{r} \log\left(\frac{c}{b \ddot{a}(T)e^{-rT}}\right) > 0 \quad \mbox{ by assumption.}\]
 Then, we can find $t'$ such that
				\begin{align*}
				& f(t)<0 \mbox{ for } t<t'\\
				& f(t)=0  \mbox{ for }  t=t'\\
				&f(t)>0  \mbox{ for }  t>t'
				\end{align*}
				Here we set $t_{*} = \lfloor{t'}\rfloor$, and we have
				\begin{align*}
				\lim_{w\rightarrow \infty} v^e(t,w) - v^h(t,w) \leq  f(t) < 0, \forall t< t_{*}
				\end{align*}
				and for $t = t_{*}$, first notice that
				\begin{align*}
				f(t_{*}) \leq 0 \implies c &\geq b \ddot{a}(T)e^{-rT}\left(t_{*}+1 - t_{*}e^{-\mu_L}\right)e^{rt_{*}}
				\end{align*}
				Next, for any finite $w>t_{*}L_{t_{*}-1}b\ddot{a}(T)e^{-r(T-t_{*})}$, we have
				\begin{align*}
				v^h(t_{*},w) &= E^Q\left[e^{-r}v\left(t_{*}+1,\left(w+cL_{t_{*}}\right)\frac{S_{t_{*}+1}}{S_{t_{*}}}\right)\right]\\
				&>E^Q\left[e^{-r}v^e\left(t_{*}+1,\left(w+cL_{t_{*}}\right)\frac{S_{t_{*}+1}}{S_{t_{*}}}\right)\right]\\ &=E^Q\left[e^{-r}\left(\left(w+cL_{t_{*}}\right)\frac{S_{t_{*}+1}}{S_{t_{*}}}-(t_{*}+1)L_{t_{*}}b \ddot{a}(T)e^{-r(T-t_{*}-1)}\right)^+\right]\\
				&\geq \max\left(0,w+cL_{t_{*}}- (t_{*}+1)L_{t_{*}}b \ddot{a}(T)e^{-r(T-t_{*})}\right)\\
				&\geq \max\left(0,w- t_{*}L_{t_{*}-1}  b \ddot{a}(T)e^{-rT}e^{rt_{*}}  \right)\\
				&= v^e(t_{*},w)
				\end{align*}
				The third to fourth line follows from Jensen's Inequality:
				\begin{align*}
	&E^Q\left[e^{-r}\left(\left(w+cL_{t_{*}}\right)\frac{S_{t_{*}+1}}{S_{t_{*}}}-(t_{*}+1)L_tb \ddot{a}(T)e^{-r(T-t_{*}-1)}\right)^+\right]\\
	&			\geq \max\left(0, E^Q\left[e^{-r}\left(\left(w+cL_{t_{*}}\right)\frac{S_{t_{*}+1}}{S_{t_{*}}}-(t_{*}+1)L_tb \ddot{a}(T)e^{-r(T-t_{*}-1)}\right)\right]\right)\\
	&			\qquad =\max\left(0, w+cL_{t_{*}} - (t_{*}+1)L_tb \ddot{a}(T)e^{-r(T-t_{*})}\right)
				\end{align*}
				
				Thus, we have $v^h(t_{*},w)>v^e(t_{*},w), \forall w<\infty$, and \[\lim_{t\rightarrow \infty} v^e(t_{*},w) - v^h(t_{*},w)\leq f(t_{*}) \leq 0\] which implies $\varphi(t_{*}) = \infty$.\\
				
Repeating the induction:
				
	\textbf{At time T}, again we have $\varphi(T) < \infty$.\\
				\textbf{At time $t>t_{*}$}, assume $\varphi(t+1)<\infty$.
				\begin{align*}
				\lim_{w\rightarrow \infty} v^e(t,w) - v^h(t,w) &= f(t) >0, t>t_{*}
				\end{align*}
				Thus $\varphi(t)<\infty, \forall t>t_{*}$.

\item[(2)] If $c>b \ddot{a}(T) \left(\left(1-e^{-\mu_L}\right)T+e^{-\mu_L}\right)e^{-r}$
			\begin{align*}
			c&> b \ddot{a}(T) e^{-r}\left(T-(T-1)e^{-\mu_L}\right)\\
			& > b \ddot{a}(T) e^{-r} \geq b \ddot{a}(T) e^{-rT}
			\end{align*}
			Thus, we know there exists $t_{*}$ as defined previously, and for time $T-1$,
			\begin{align*}
			\lim_{w\rightarrow \infty}&v^e(T-1,w) - v^h(T-1,w)\\
			&= TbL_{T-1} \ddot{a}(T)e^{-r}-(T-1)bL_{T-2} \ddot{a}(T)e^{-r}-cL_{T-1}\\
			&< TbL_{T-1} \ddot{a}(T)e^{-r}-(T-1)bL_{T-2} \ddot{a}(T)e^{-r}-b \ddot{a}(T) \left(\left(1-e^{-\mu_L}\right)T+e^{-\mu_L}\right)e^{-r}L_{T-1}\\
			&= 0
			\end{align*}
			Thus, $\varphi(t) = \infty, \forall t\leq T-1$, and the option simplifies to the DB underpin option.
		
			\item[(3)] 	  When $\displaystyle{\frac{c}{b \ddot{a}(T)e^{-rT}}=1}$, we have $f(0) = 0$. Thus $t_{*}=t' = 0$, immediately we have $\varphi(0) = \infty$.\\
			
		\end{enumerate}

	\subsection{Formulation of Value Function - Continuous Case\label{appE}}
	Recall $S_t$, $L_t$ represent the stock process and the salary process, and, $W_t$ represents the wealth accumulation process in the DC account. Their stochastic differential equation representations are:
	\begin{align*}
	dS_t &= rS_tdt + \sigma_SS_tdZ_S^Q(t)\\
	dL_t &= rL_tdt + \sigma_LL_tdZ_L^Q(t)\\
	dW_t &= rW_tdt + cL_tdt + \sigma_SW_tdZ_S^Q(t)\\
	\rho dt &= dZ_L^Q(t) dZ_S^Q(t)
	\end{align*}
	Here we denote $Y_t = \frac{W_t}{L_t}$, as the wealth-salary ratio process. We can rewrite our value function:
	
		\begin{align*}
		v(t,w, l)&=\sup_{0\leq \tau \leq T-t}E^Q\left[ e^{-r\tau }\left(W_{\tau+t}^{t,w}-K_{t+\tau}e^{-r(T-t-\tau)}\right)^+|{\mathcal{F}}_t \right]\\
		&=\sup_{0\leq \tau \leq T-t}E^Q\left[ e^{-r\tau }L_{\tau+t}\left(\frac{w}{L_{\tau+t}}\frac{S_{t+\tau}}{S_t} + c\int_{0}^{\tau}\frac{S_{\tau+t}L_u}{S_uL_{t+\tau}}-b(t+\tau)\ddot{a}_Te^{-r(T-t-\tau)}\right)^+\bigg|{\mathcal{F}}_t \right]\\
		&=\sup_{0\leq \tau \leq T-t}E^Q\left[ e^{-r\tau}L_{\tau+t}\left(Y_{t+\tau}^{t,w/l}-b(t+\tau)\ddot{a}_Te^{-r(T-t-\tau)}\right)^+\bigg|{\mathcal{F}}_t \right]\\
		&\sup_{0\leq \tau \leq T-t}E^Q\left[ e^{-r\tau}le^{\left(r-\frac{\sigma_L^2}{2}\right)\tau + \sigma_L(Z_L^Q(t+\tau)-Z_L^Q(t))}\left(Y_{t+\tau}^{t,w/l}-b(t+\tau)\ddot{a}_Te^{-r(T-t-\tau)}\right)^+\bigg|{\mathcal{F}}_t \right]\\
		\end{align*}
		We are able to eliminate the discounting term in the expectation through the change of measure method. Let $
		d\tilde{P} = \exp\left(\sigma_L Z^Q_L(t) - (\sigma_L^2/2)t\right)dQ$, by Girsanov Theorem, we have
		\begin{align*}
		\begin{pmatrix}
		Z_S^{\tilde{P}}(t)\\
		Z_L^{\tilde{P}}(t)
		\end{pmatrix} = \begin{pmatrix}
		Z_S^Q\\
		Z_L^Q
		\end{pmatrix} - \begin{pmatrix}
		0\\\sigma_L
		\end{pmatrix}t
		\end{align*}
		as a two-dimensional standard Brownian motion under $\tilde{P}$. Then, under the new measure, $(S_t, L_t)$ has the following SDE:
		\begin{align*}
		\begin{pmatrix}
		dS_t\\
		dL_t\\
		\end{pmatrix} &= \begin{pmatrix}
		rS_t + \sigma_S\sigma_L\rho S_t\\
		rL_t + \sigma_L^2 L_t\\
		\end{pmatrix}dt + \begin{pmatrix}
		\sigma_S S_t \sqrt{1-\rho^2} & \sigma_S S_t \rho\\0 & \sigma_L L_t
		\end{pmatrix}
		\begin{pmatrix}
		dZ_S^{\tilde{P}}(t)\\dZ_L^{\tilde{P}}(t)
		\end{pmatrix}
		\end{align*}
		and the SDE for the wealth process is
		\begin{align*}
		dW_t &= (rW_t + \sigma_S\sigma_L\rho W_t + cL_t)dt + \sigma_S W_t \sqrt{1-\rho^2}dZ_S^{\tilde{P}}(t) + \sigma_SW_t\rho dZ_L^{\tilde{P}}(t)
		\end{align*}
		
		then our value function becomes
		\begin{align*}
		v(t,w,l)=&\sup_{0\leq \tau \leq T-t}E^Q\left[ e^{-r\tau}le^{\left(r-\frac{\sigma_L^2}{2}\right)\tau + \sigma_L(Z_L^Q(t+\tau)-Z_L^Q(t))}\left(Y_{t+\tau}^{t,w/l}-b(t+\tau)\ddot{a}_Te^{-r(T-t-\tau)}\right)^+\bigg|{\mathcal{F}}_t \right]\\
		=&\sup_{0\leq \tau \leq T-t}E^{\tilde{P}}\left[\frac{dQ}{d\tilde{P}}le^{-\frac{\sigma_L^2}{2}\tau + \sigma_L(Z_L^Q(t+\tau)-Z_L^Q(t))}\left(Y_{t+\tau}^{t,w/l}-b(t+\tau)\ddot{a}_Te^{-r(T-t-\tau)}\right)^+\bigg|{\mathcal{F}}_t \right]\\
		=& l\sup_{0\leq \tau \leq T-t}E^{\tilde{P}}\left[\left(Y_{\tau+t}^{t,w/l}-b(\tau+t)\ddot{a}_Te^{-r(T-t-\tau)}\right)^+\bigg|{\mathcal{F}}_t \right]
		\end{align*}
		where $Y_t$ has the SDE as:
		\begin{align*}
		dY_t &= \frac{1}{L_t}dW_t - \frac{W_t}{L_t^2}dL_t -\frac{1}{L_t^2}dL_tdW_t + \frac{W_t}{L_t^3}\left(dL_t\right)^2\\
		&= \left(r+\sigma_S\sigma_L\rho\right) Y_t dt + cdt + \sigma_S\sqrt{1-\rho^2}Y_tdZ_S^{\tilde{P}}(t) + \sigma_S\rho Y_t dZ_L^{\tilde{P}}(t)\\
		& \qquad - Y_t \left((r+\sigma^2_L)dt + \sigma_L dZ_L^{\tilde{P}}(t)\right) - Y_t\sigma_L\sigma_S\rho dt + Y_t \sigma_L^2 dt\\
		&=  cdt + \sigma_S\sqrt{1-\rho^2}Y_tdZ_S^{\tilde{P}}(t) + Y_t\left(\sigma_S\rho-\sigma_L\right)dZ_L^{\tilde{P}}(t)\\
		&=  cdt + Y_t \sigma_YdZ_Y^{\tilde{P}}(t)
		\end{align*}
		where $\sigma_Y = \sqrt{\sigma_S^2 + \sigma_L^2 - 2\sigma_S\sigma_L\rho}$ and $Z_Y^{\tilde{P}}(t)$ is a standard Brownian Motion. We can define a new function (2-dimensional):
		\begin{align*}
		v(t,y) = \sup_{0\leq \tau \leq T-t}E^{\tilde{P}}\left[\left(Y_{t+\tau}^{t,y}-b(t+\tau)\ddot{a}_Te^{-(T-t-\tau)r}\right)^+\bigg|{\mathcal{F}}_t \right]
		\end{align*}
		Thus, $v(t,w/l) = v(t,w,l)/l$, which clearly suggests that the exercise rule depends on the wealth-to-salary ratio.
		
		\subsection{Characteristics of Value Function - Continuous Case\label{appF}}
			
		\subsubsection{Non-decreasing in y}
		By writing value function explicitly
		\begin{align*}
		v(t,y) = \sup_{0\leq \tau \leq T-t}E^{\tilde{P}}\left[\left(y\frac{S(t+\tau)}{L(t+\tau)S(t)}+c\int_{t}^{t+\tau}\frac{S(t+\tau)L(u)}{S(u)L(t+\tau)}du-b(t+\tau)\ddot{a}_Te^{-(T-t-\tau)r}\right)^+\right]
		\end{align*}
		we immediately see that $y\rightarrow v(t,y)$ is an increasing and convex function on $[0, \infty)$

		\subsubsection{Continuity of value function in y}
		for $x>y$
		\begin{align*}
	&	|v(t,x) - v(t,y)|\\
&\leq \Bigg|\sup_{0\leq \tau \leq T-t}E^{\tilde{P}}\left[\left(Y_{t+\tau}^{t,x}-b(t+\tau)
\ddot{a}_Te^{-(T-t-\tau)r}\right)^+-\left(Y_{t+\tau}^{t,y}-b(t+\tau)\ddot{a}_Te^{-(T-t-\tau)r}\right)^+
\bigg|{\mathcal{F}}_t \right]\Bigg|\\
		&\leq \left| \sup_{0\leq \tau \leq T-t}E^{\tilde{P}}\left[Y_{t+\tau}^{t,x}-Y_{t+\tau}^{t,y} \right]  \right|\\
		&\leq \left| \sup_{0\leq \tau \leq T-t}E^{\tilde{P}}\left[x\frac{L_tS_{t+\tau}}{L_{t+\tau}S_t}-y\frac{L_tS_{t+\tau}}{L_{t+\tau}S_t}\right]  \right|\\
		&\leq \left| (x-y) \sup_{0\leq \tau \leq T-t}E^{\tilde{P}}\left[\frac{L_tS_{t+\tau}}{L_{t+\tau}S_t}\right]  \right|\\
		&\leq (x-y)
		\end{align*}
				
		\subsubsection{Non-increasing in time t}
		Since $Y_{t_1+t}^{t_1,y} \stackrel{law}{=} Y_{t_2+t}^{t_2,y}$, and the strike function $K(t) = tbe^{-(T-t)r}\ddot{a}_T$ is an increasing function of time t, immediately we have $v(t,y)$ is non-increasing in t.

		\subsubsection{Continuity of value function in t}
		For $t_2>t_1$,
		\begin{align*}
		0\leq &v(t_1,y) - E^{\tilde{P}}\left[v(t_2,Y_{t_2}^{t_1,y})\right]\\
		=& \sup_{\tau_1\leq T-t_1}E^{\tilde{P}}\left[\mathbf{1}_{\tau_1<t_2-t_1}v^e\left(t_1+\tau_1,Y_{t_1+\tau_1}^{t_1,y}\right)+\mathbf{1}_{\tau_1\geq t_2-t_1}v\left(t_2,Y_{t_2}^{t_1,y}\right)\right]- E^{\tilde{P}}\left[v(t_2,Y_{t_2}^{t_1,y})\right]\\
		\leq&\sup_{\tau_1\leq T-t_1}E^{\tilde{P}}\left[\mathbf{1}_{\tau_1<t_2-t_1}v^e\left(t_1+\tau_1,Y_{t_1+\tau_1}^{t_1,y}\right)+\mathbf{1}_{\tau_1\geq t_2-t_1}v\left(t_2,Y_{t_2}^{t_1,y}\right)-v(t_2,Y_{t_2}^{t_1,y})\right]\\
		=&\sup_{\tau_1\leq T-t_1}E^{\tilde{P}}\left[\mathbf{1}_{\tau_1<t_2-t_1}v^e(t_1+\tau_1,Y_{t_1\tau_1}^{t_1,y})-v(t_2,Y_{t_2}^{t_1,y})\right]\\
		\leq&\sup_{\tau_1\leq T-t_1} E^{\tilde{P}}\left[\mathbf{1}_{\tau_1<t_2-t_1}v^e(t_1+\tau_1,Y_{t_1\tau_1}^{t_1,y})-v^e(t_2,Y_{t_2}^{t_1,y})\right]\\
		\leq&\sup_{\tau_1\leq T-t_1} E^{\tilde{P}}\left[\mathbf{1}_{\tau_1<t_2-t_1}\left(Y_{\tau_1+t_1}^{t_1,y} - (t_1+\tau_1)b\ddot{a}_Te^{-(T-t_1-\tau_1)r}\right)^+-\left(Y_{t_2}^{t_1,y} - t_2b\ddot{a}_Te^{-(T-t_2)r}\right)^+\right]\\
		\leq & E^{\tilde{P}}\left[\sup_{\tau_1\in[t_1,t_2]}\left|Y_{t_1+\tau_1}^{t_1,y} - Y_{t_2}^{t_1,y}\right| + \sup_{\tau_1\in [t_1,t_2]}\left|t_2b\ddot{a}_Te^{-(T-t_2)r}-(t_1+\tau_1)b\ddot{a}_Te^{-(T-t_1-\tau_1)r}\right|\right]\\
		\leq &E^{\tilde{P}}\left[\sup_{\tau_1\in[t_1,t_2]}\left|Y_{t_1+\tau_1}^{t_1,y} - Y_{t_2}^{t_1,y}\right| \right] + (t_2b\ddot{a}_Te^{-(T-t_2)r}-t_1b\ddot{a}_Te^{-(T-t_1)r})\\
		\leq &C_1 \sqrt{t_2-t_1}
		\end{align*}
		where the last line comes from \citet{touzi}, Theorem 2.4, that there exists a constant $C$ such that
		\begin{align*}
		E^{\tilde{P}}\left[\sup_{\tau_1\in[t_1,t_2]}\left|Y_{t_1+\tau_1}^{t_1,y} - Y_{t_2}^{t_1,y}\right| \right] \leq C (1+|y|)\sqrt{t_2-t_1}
		\end{align*}
		Next, we have
		\begin{align*}
		|v(t_1,y)-v(t_2,y)| &\leq \left|v(t_1,y) - E^{\tilde{P}}\left[v(t_2,Y_{t_2}^{t_1,y})\right] \right| + \left|E^{\tilde{P}}\left[v(t_2,Y_{t_2}^{t_1,y})\right]-v(t_2,y)\right|\\
		&\leq C_1 \sqrt{t_2-t_1} + E^{\tilde{P}}\left[\left|v(t_2,Y_{t_2}^{t_1,y})-v(t_2,y)\right|\right]\\
		&\leq C_1 \sqrt{t_2-t_1} + C_2 E^{\tilde{P}}\left[\left|Y_{t_2}^{t_1,y}-y\right|\right]\\
		&\leq C \sqrt{t_2-t_1}
		\end{align*}
		where the third line follows from the Lipschitz continuity of the value function in y. Thus, we have proved that the value function is H\"older-Continuous in t.\\

		\subsubsection{Exercise Region}
		Denote the continuation region $\mathbf{C} = \left\{(t,y)\in (0,T)\times [0,\infty): v(t,y)>v^e(t,y) \right\}$ and stopping regions $\mathbf{C} = \left\{(t,y)\in (0,T)\times [0,\infty): v(t,y)=v^e(t,y) \right\}$. Then, by standard arguments based on the strong Markov Property (see \citet{peskir} Corollary 2.9), the first hitting time $\tau_{D} = \inf\left\{0\leq s \leq T-t: (t+s, Y_{t+s})\in D\right\}$ is optimal, and the value function is $C^{1,2}$ on $\mathbf{C}$ and satisfies:
		\begin{align*}
		v_t + \mathbf{L}_Yv = 0 \quad in \quad \mathbf{C}
		\end{align*}
		
		If $(t,y) \in \mathbf{D}$,
		\begin{align*}
		v(t,y+h)\leq v(t,y) + h = v^e(t,y) + h = v^e(t,y+h)
		\end{align*}
		thus $(t,y+h) \in \mathbf{D}$.\\
			
		If $(t,y) \in \mathbf{C}$,
		\begin{align*}
		v(t,y-h)\geq v(t,y) -h > v^e(t,y) - h = v^e(t,y-h), \quad y-h>K_t
		\end{align*}
		if $y-h<K_t$, $v(t,y-h)>0 = v^e(t,y-h)$, thus $(t,y-h)\in \mathbf{C}$\\

		\subsubsection{Conditions when $\varphi(t) = \infty$}
		Define and substitute $g(t,y) = (y-bt\ddot{a}_Te^{-r(T-t)})$ for $y>K_t$ into the variational inequality, and define
		\begin{align*}
		f(t) = -\frac{\partial g}{\partial t}-c\frac{\partial g}{\partial y} - \frac{\sigma_Y^2y^2}{2}\frac{\partial^2 g}{\partial y^2} = b\ddot{a}_Te^{-r(T-t)} + rbt\ddot{a}_Te^{-r(T-t)} - c
		\end{align*}
		If $f(t)<0$, then there is a contradiction with the variational inequality, which means $v(t,y) \neq g(t,y)$ for all y, and thus $(t,y)\in \mathbf{C}, \forall y$. Since $f(t)$ is an increasing function of t, there exists a $t^*$ satisfying $f(t^*) = 0$. If $t^{*}\in[0,T]$, then $(t,y)\in \mathbf{C}, \forall (t,y)\in [0,t^*]\times \mathbf{R}$. In particular, if $c>b\ddot{a}_T(1+rT)$ (when DC contribution rate is extremely high or horizon is short), the option is equivalent to European Option.

		\subsection{Price of Second Election}\label{appG}
		Here we provide the pricing formulae for second election option under three scenarios.
		\begin{itemize}
			\item Stochastic Salary in Continuous Setting:
			\begin{align*}
			C^s(t,L_t)=&\left(L_tc(t^{*}-t)-t^{*} L_tb\ddot{a}_Te^{-r(T-t^{*})} \right) + Tb\ddot{a}_TL_t
			\end{align*}
			which we can find a $t^{*} = \max(\min(T,t^{'}),t)$, that $t^{'}$ satisfies
			\begin{align*}
			c - b\ddot{a}_Te^{-r(T-t^{'})}- rt^{'}b\ddot{a}_Te^{-r(T-t^{'})} = 0
			\end{align*}
			\item Deterministic Salary in Continuous Setting:
			\begin{align*}
			C^s(t,L_t)=\frac{cL_t}{\mu_L-r} \left(e^{(\mu-r)(t^{*}-t)}-1\right)-t^{*}b\ddot{a}_TL_te^{(\mu_L-r)(t^{*}-t)}e^{-\gamma(T-t^{*})} + TL_te^{(\mu_L-r)(T-t)}b\ddot{a}_T
			\end{align*}
			where $t^{*} = \min(\max(t,t^{'}),T)$ and $t^{'}$ satisfies
			\begin{align*}
			c-b\ddot{a}_Te^{-\gamma(T-t^{'})} - t^{'}(\mu_L-r+\gamma)b\ddot{a}_Te^{-\gamma(T-t^{'})}= 0
			\end{align*}
			\item Deterministic Salary in Discrete Setting:
			\begin{align*}
&			C^s(t,L_t)\\
&=\max_{t^{*}\in t,t+1,\cdots T} L_t\left(c\frac{1-e^{(\mu_L-r)(t^{*}-t)}}{1-e^{\mu_L-r}}-t^{*}b\ddot{a}_Te^{\mu_L(t^{*}-1-t)}e^{-rt^{*}}e^{-\gamma(T-t^{*})} + Tb\ddot{a}_Te^{\mu_L(T-1)}e^{-r(T-t)}\right)
			\end{align*}
		\end{itemize}

\end{document}